\documentclass[11pt,a4paper]{article}
\usepackage[utf8]{inputenc}
\usepackage{amsmath}
\usepackage{amsfonts}
\usepackage{amssymb}
\usepackage{amsthm}
\usepackage{makeidx}
\usepackage{graphicx}
\usepackage[left=2.00cm, right=1.50cm, top=3.00cm, bottom=3.00cm]{geometry}
\usepackage{url}
\usepackage[square,sort,comma,numbers]{natbib}

\newtheorem{theorem}{Theorem}[section]

\begin{document}

\begin{center}
\huge{\textbf{A Differentiable Alternative to the Lasso Penalty}}
\end{center}
\normalsize

\begin{center}
{Hamed Haselimashhadi, Veronica Vinciotti}
\end{center}

\begin{abstract}
Regularized regression has become very popular nowadays, particularly on high-dimensional problems where the addition of a penalty term to the log-likelihood allows inference where traditional methods fail. A number of penalties have been proposed in the literature, such as lasso, SCAD, ridge and elastic net to name a few. Despite their advantages and remarkable performance in rather extreme settings, where $p \gg n$, all these penalties, with the exception of ridge, are non-differentiable at zero. This can be a limitation in certain cases, such as computational efficiency of parameter estimation in non-linear models or derivation of estimators of the degrees of freedom for model selection criteria. With this paper, we provide the scientific community with a differentiable penalty, which can be used in any situation, but particularly where differentiability plays a key role. We show some desirable features of this function and prove theoretical properties of the resulting estimators within a regularized regression context. A simulation study and the analysis of a real dataset show overall a good performance under different scenarios. The method is implemented in the R package {\tt DLASSO} freely available from CRAN, \url{http://CRAN.R-project.org/package=DLASSO}.
\end{abstract}


\section{Introduction}
In the usual regularized regression situation, the regression parameters $\beta=(\beta_1\ldots\beta_p)$ are estimated by minimising
\begin{equation}
 \sum_{i=1}^n(y_i-x_i\beta')^2 + \lambda p(\beta),
 \label{eq:penreg}
\end{equation}
with $y=(y_1,\ldots,y_n)$ the observations on the response $y$, which we assume to be centered, and $x_i=(x_{i1}\ldots x_{ip})$, $i=1,\ldots,n$, the observations on the $p$ covariates. Various forms of the penalty function $p(\beta)$ have been suggested in the literature, such as the $l_2$ norm, as in ridge regression \citep{hoerl70}, the $l_1$ norm as in the popular lasso regression \citep{tibshirani96}, the Smoothly Clipped Absolute Deviation (SCAD) penalty \citep{fan01}, the fused lasso \citep{tibshirani05}, group lasso \citep{yuan06}, combinations of $l_1$ and $l_2$ norms, such as elastic net \citep{zou05} and the Smooth-Lasso \citep{hebiri11}. Aside from the ridge penalty, which does not necessarily lead to sparsity and variable selection, all of the other penalties are non-differentiable at zero.  This can be a limitation in certain cases, such as computational efficiency for non-linear models \citep{schmidt07} or derivation of the degrees of freedom for model selection criteria, such as the generalised information criterion \citep{konishi96}, as pointed out by \cite{abbruzzo14}.

In this paper, we address this gap by proposing a penalty function that is differentiable at zero, and which possesses also many of the desirable properties of existing penalty functions. The function has one tuning parameter, by varying, which one can obtain a penalty extremely close to the absolute value (lasso) or a quadratic function (ridge) or combinations of these. In Section \ref{sec:dlasso}, we define this new penalty function, which we call {\tt dlasso}, and list its properties. In Section \ref{sec:dlasso-reg}, we study the properties of the estimators under a {\tt dlasso} penalty in a regularized regression context. In Section \ref{sec:dlasso-alg}, we provide an efficient algorithm for parameter estimation in regularized regression, by exploiting the differentiability of the penalty function. In Section \ref{sec:simulation} and \ref{sec:realdata}, we study the performance of this new approach on a number of simulated scenarios and on a real dataset, by comparing it with existing methods. Finally, in Section \ref{sec:conclusion}, we draw some conclusions and point to directions for future work.

\section{Our proposal: dlasso} \label{sec:dlasso}
Looking at the literature for differentiable approximations of the absolute value, a number of proposals have been made, such as
\begin{align}
& |x| \approx  \sqrt{x^2+s} ,\qquad s\in \mathbb{R}_+, \text{\citep{ramirez14}}\label{appAbs:1} \\
& |x| \leq \sqrt{x^2+s^2}, \quad s\in \mathbb{R}_+, \text{\citep{nesterov05}} \label{appAbs:2} \\
& |x| \geq \frac{x^2}{\sqrt{x^2+s^2}}, \quad s\in \mathbb{R}_+, \text{\citep{nesterov05}} \label{appAbs:3} \\
& |x| \approx s\log(2+e^{-x/s}+e^{x/s}), \quad s\in \mathbb{R}_+, \text{\citep{schmidt07}}. \label{appAbs:4}
\end{align}
Equation (\ref{appAbs:1}) is a special case of (\ref{appAbs:2}) and the length of the interval from equation (\ref{appAbs:2}) and (\ref{appAbs:3}) is always less than $u$~\citep{nesterov05}. The approximation (\ref{appAbs:4}) has been used by \cite{schmidt07} in a penalized likelihood context. This function is twice differentiable and $|x|=\lim\limits_{s \rightarrow 0}f(x)$ with the maximum absolute difference of  $\big\vert |x|-f(x)\big\vert \leq 2s\log(2)$, but it does not pass through zero. This, however, is a desirable property for a penalty function if one wants the tuning parameter to cover a number of penalties such as $l_2$.

Motivated by this challenge, and noting some advantageous properties of the error function \citep{olver10}, in this paper we propose the following penalty function
\begin{equation}
p(x,s)=x\Big(\frac{2}{\sqrt{\pi}} \int_{0}^{x/s}e^{-t^2}dt\Big), \quad s \in \mathbb{R}_+. \label{appAbs:5}
\end{equation}
The function can be written in different forms,
\[p(x,s)=x\rm{erf}\Big(\frac{x}{s}\Big)=x\Big(1-\rm{erf}^c\Big(\frac{x}{s}\Big)\Big)=x\Big(2\Phi\Big(\frac{x}{s},0,\dfrac{1}{\sqrt{2}}\Big)-1\Big),\]
either in terms of the error function {\tt erf}$=\frac{2}{\sqrt{\pi}} \int_{0}^{x/s}e^{-t^2}dt$, or of its complementary {\tt erf$^c$} = 1-erf, or of the cdf of a normal distribution with mean 0 and standard deviation $\dfrac{1}{\sqrt{2}}$, which we denote by $\Phi\Big(x,0,\dfrac{1}{\sqrt 2}\big)$.

The function has a number of properties, some of which make it an appealing choice for regularized inference:
\begin{enumerate}
\item $p(0,s)=0$ for any $s$.
\item $p(x,s)$ is twice differentiable with respect to $x$, with the derivatives given by
\begin{align*}
\frac{d}{dx}p(x,s)&=\rm{erf}\Big(\frac{x}{s}\Big)+2\text{$\phi$}\Big(\frac{x}{s},0,\frac{1}{\sqrt{2}}\Big)\frac{x}{s},\\
\frac{d^2}{dx}p(x,s) &= \frac{2}{s} \phi\Big(\frac{x}{s},0,\frac{1}{\sqrt{2}}\Big) + \frac{2}{s} \phi\Big(\frac{x}{s},0,\frac{1}{\sqrt{2}}\Big) - \frac{4}{x}\Big(\frac{x}{s}\Big)^{3} \phi\Big(\frac{x}{s},0,\frac{1}{\sqrt{2}}\Big)=4\phi\Big(\frac{x}{s},0,\frac{1}{\sqrt{2}}\Big)\frac{1}{s}\bigg(1-\Big(\frac{x}{s}\Big)^{2}\bigg),
\end{align*}
with $\phi$ the density function of the normal distribution. Note that, similarly to the SCAD penalty, the dlasso penalty is not convex. For example the second derivative is positive if  $\bigg(1-\Big(\dfrac{x}{s}\Big)^{2}\bigg)>0$ or $|x|<s$.
\item As $s \rightarrow 0$, the function converges to $|x|$ exponentially fast. In fact, we prove that \[\big\vert|x|-p(x,s) \big\vert \leq 2s\phi\Big(\frac{x}{s},0,\frac{1}{\sqrt{2}}\Big)\:\: \text{for all}\: x \: \text{and for}\: s>0.\]
\begin{proof}
For the proof, we use the bound on the complementary error function given by ~\cite{abramowitz12}
\[\frac{2}{\sqrt{\pi}}\frac{e^{-(\frac{x}{s})^2}}{(\frac{x}{s})+\sqrt{(\frac{x}{s})^2+2}} <\rm{erf}^c\Big(\frac{x}{s}\Big) \leq \frac{2}{\sqrt{\text{$\pi$}}}\frac{e^{-(\frac{x}{s})^2}}{(\frac{x}{s})+\sqrt{(\frac{x}{s})^2+\frac{4}{\text{$\pi$}}}}.\]
Using the inequalities above we get,
\begin{align*}
x>0 \rightarrow &\big\vert x-x\Big(1-\rm{erf}^c\Big(\frac{x}{s}\Big)\Big) \big\vert = \big\vert x\rm{erf}^c\Big(\frac{x}{s}\Big) \big\vert
\leq
\frac{2x}{\sqrt{\text{$\pi$}}}\frac{e^{-(\frac{x}{s})^2}}{(\frac{x}{s})+\sqrt{(\frac{x}{s})^2+\frac{4}{\text{$\pi$}}}}\\
&=\frac{2s}{\sqrt{\pi}}e^{-(\frac{x}{s})^2}\frac{1}{1+\sqrt{1+\frac{4s^2}{\pi x^2}}} \leq \frac{2s}{\sqrt{\pi}}e^{-(\frac{x}{s})^2}={{2}s}\phi(\frac{x}{s},0,\frac{1}{\sqrt{2}}).
\end{align*}
Following a similar approach for $x<0$ leads to the same result. Consequently,
\[\big\vert|x|-p(x,s) \big\vert \leq 2s{{}}\phi(\frac{x}{s},0,\frac{1}{\sqrt{2}}).\]
\end{proof}
The right hand side (RHS) of the inequality tends to zero as $s\rightarrow 0$ at an exponential speed. Figure 1 (left) accompanies this result, by showing that our chosen function converges to the absolute value faster than its opponents.

\begin{figure}[ht!]
\centering
\includegraphics[scale=0.5]{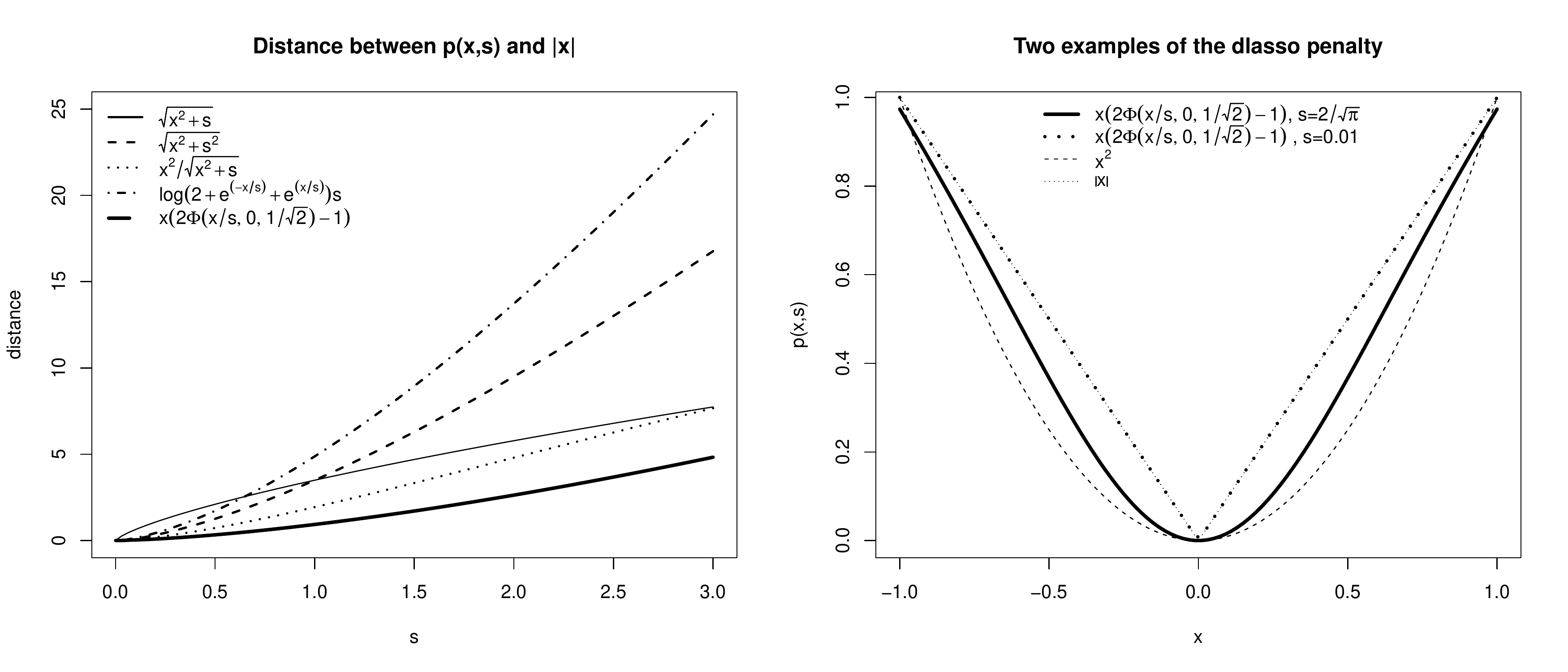}
\caption{Key properties of the dlasso penalty. Left: Fast convergence to the absolute value compared to other approximations.  Right: dlasso is like lasso for small $s$ and like ridge for $s=\frac{2}{\sqrt{\pi}}$ and small $x$.}
\label{fig:speedconv}
\end{figure}

\begin{center}
FIGURE 1 ABOUT HERE
\end{center}

\item If we set $s=\dfrac{2}{\sqrt{\pi}}$, the function behaves like the $l_2$ norm in the vicinity of zero. \\
This is due to the fact that for small $x$:
\[\frac{2}{\sqrt{\pi}} x\int_{0}^{x/s}e^{-t^2}dt \approx \frac{2}{\sqrt{\pi}} \frac{x^2}{s}e^{-(\frac{x}{s})^2}=x^2 \frac{2}{s}\phi(\frac{x}{s},0,\frac{1}{\sqrt{2}}).\]
Then, setting $s=\dfrac{2}{\sqrt{\pi}}$, the RHS becomes $\sqrt{\pi}x^2 ~\phi(\frac{\sqrt{\pi}x}{2},0,\frac{1}{\sqrt{2}})$. On the other hand, $\phi(\frac{\sqrt{\pi}x}{2},0,\frac{1}{\sqrt{2}})\overset{x\rightarrow 0}{\approx} \frac{1}{\sqrt{\pi}}$, from which $\sqrt{\pi}x^2 \phi(\frac{\sqrt{\pi}x}{2},0,\frac{1}{\sqrt{2}})\approx x^2$.
\end{enumerate}
The last two points are summarized in Figure 1 (right): when $s$ is small the function behaves like the absolute value, when $s=\dfrac{2}{\sqrt{\pi}} \approx 1$, the function behaves like $x^2$ in the vicinity of $x=0$.

The final point to discuss is about computational complexity, which is the only potential difficulty with our proposal. However, a number of good and fast approximations are provided in the literature for evaluating the error function or the cdf of a normal distribution. One option is to use Taylor approximations. For instance, approximations can be based on one of the expressions below
\begin{align}
\rm{erf}(x)&=\frac{2x}{\sqrt{\pi}}\sum_{j=0}^{\infty}\frac{(-1)^jx^{2j}}{j!(2j+1)} \label{erf:AP.eq1}\\
&=\frac{2xe^{-x^2}}{\sqrt{\pi}}\sum_{j=0}^{\infty}\frac{2^j x^{2j}}{1\cdot 3\cdot 5\, \cdots (2j+1)}\label{erf:AP.eq2},\\
\rm{erf}^c(x)&\approx\frac{e^{-x^2}}{x\sqrt{\text{$\pi$}}}\sum_{j=0}^{k}(-1)^j \frac{(2j)!}{j!}(2x)^{-2j}. \label{erf:AP.eq3}
\end{align}
For small $|x|$, the series in (\ref{erf:AP.eq1}) is slightly faster than the series in (\ref{erf:AP.eq2}) because there is no need to compute an exponential. However, the series (\ref{erf:AP.eq2}) is preferable to (\ref{erf:AP.eq1}) for moderate $|x|$ because it involves no cancellation. For large $|x|$, neither series are satisfactory and in this case it is preferable to use the asymptotic expansion for the complementary error function (\ref{erf:AP.eq3}).

Beside Taylor approximations, there are alternative fast  algorithms that approximate the error function or the normal cdf, see for example~\citep{vazquez12,olver10,chevillard08,press92,lee92,cody90,borjesson79}. In particular two fast approximations are given by
\begin{align*}
\rm{erf}(x)&\approx \tanh\bigg(\frac{39x}{2\sqrt{\text{$\pi$}}}-\frac{111}{2}\arctan(\frac{35x}{111\sqrt{\text{$\pi$}}})\bigg),\\
\Phi(x,0,1) &\approx \bigg(\frac{1}{1.9\sqrt{\pi}} \bigg(\sin(\frac{\pi x}{10})+\sin(x) \bigg)+.5\bigg)I_{(|x|\leq 1.513859)}\\
&+ \bigg( 1-e^{-1.78}+\frac{x}{e^{x+10}}\bigg)I_{(x>1.513859)}+ \bigg( e^{-1.78|x|}-\frac{|x|}{e^{|x|+10}}\bigg)I_{(x<-1.513859)}.
\end{align*}
These are not only fast but also very precise, e.g. the maximum absolute error of the last function is  $10^{-4}$.

\section{Regularized regression based on dlasso} \label{sec:dlasso-reg}
In this section, we use the new penalty function in the traditional regularized regression context and discuss the theoretical properties of the resulting estimators.

Using the new penalty function in equation (\ref{eq:penreg}) results in the optimization problem
\begin{align} L(\beta)=(y-X\beta)'(y-X\beta)+ \lambda \sum_{j=1}^{p} \beta_i \bigg( 2\Phi(\frac{\beta_i}{s},0,\frac{1}{\sqrt{2}})-1 \bigg)\quad  s>0,\lambda \geq 0.
\label{appAbs:10}
\end{align}
 In order to get an insight into the resulting estimators, let us consider the case $y_i=\beta_i+\epsilon_i$, $i=1,\ldots,n$ and $\epsilon_i \sim N(0,\sigma^2)$. Then $\hat \beta_i$ are the solutions of
\begin{align*}
\frac{d}{d\beta_i}L(\beta)=\lambda\bigg(2\Phi(\frac{\beta_i}{s},0,\frac{1}{\sqrt{2}})-1+2(\frac{\beta_i}{s}) \phi(\frac{\beta_i}{s},0,\frac{1}{\sqrt{2}})\bigg)-2(y_i-\beta_i)=0, \: i=1,\ldots,n.
%\label{ABS:General:2}
\end{align*}
Figure 2 %(\ref{fig:thresholding})
shows the estimators for a range of values of $s$. As discussed before and as evident from this plot, (a), (c) and (d) show similar regularizations to lasso, ridge and non-penalized linear regression, respectively.

\begin{center}
FIGURE 2 ABOUT HERE
\end{center}

\begin{figure}[ht!]
\centering
\includegraphics[scale=0.7]{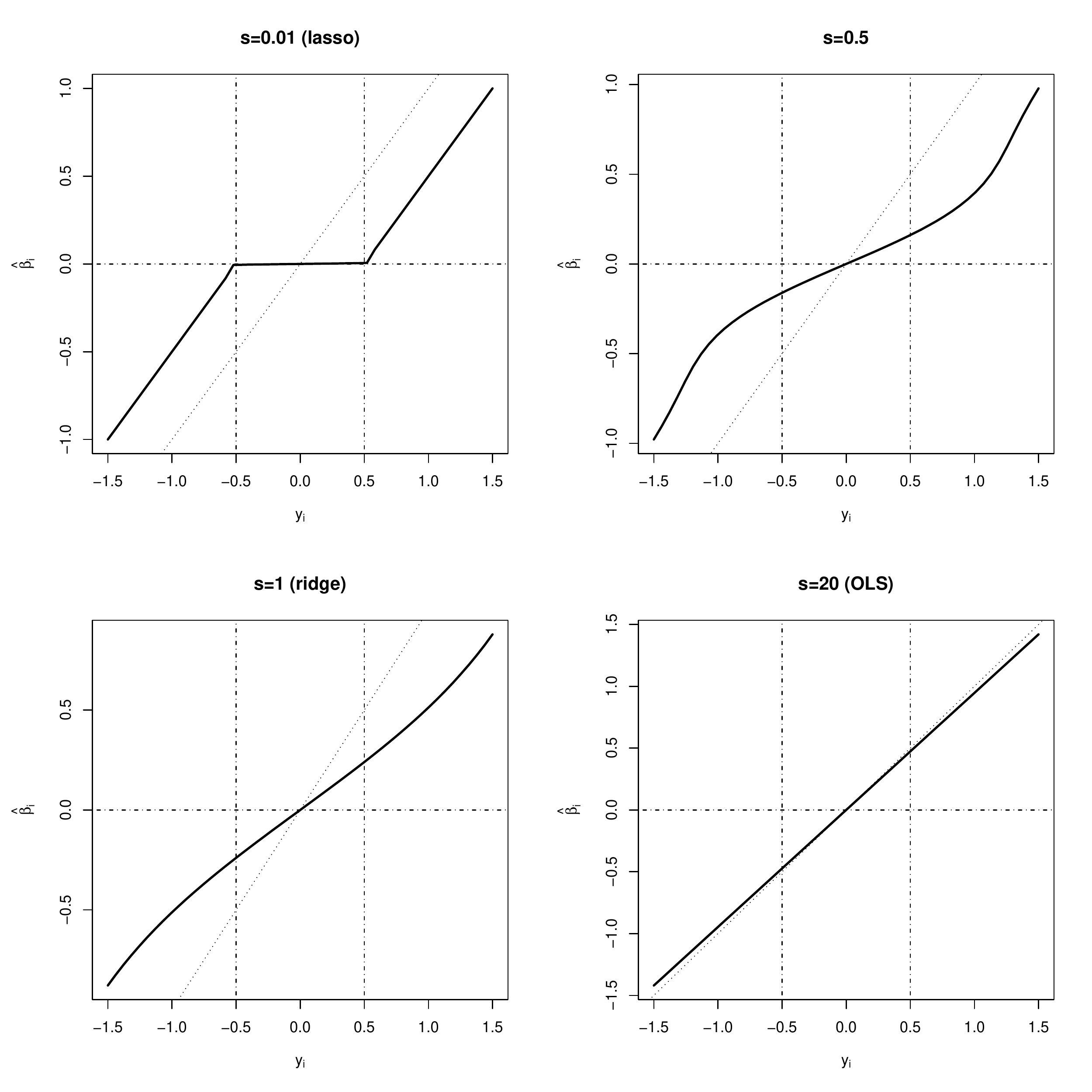}
\caption{Plot of thresholding functions with $\lambda=1$ and $s$ assuming different values: $s=0.01$ (lasso), $s=0.5$, $s=1$ (ridge), $s=20$ (ordinary least squares).}
\label{fig:thresholding}
\end{figure}

Given the very good approximation of the function to the absolute value when $s$ is close to zero, we expect the estimators to have similar properties to the lasso estimators in this case. This is proved by the next two theorems, where we follow a similar approach to~\citep{knight00}.
\begin{theorem}\label{ch2:theorem1}
For any $u\in \mathbb{R}^p$, $\lambda\geq 0$ and $s>0$ define,
\begin{align*}
k(u,s) = L(\beta+u)-L(\beta),
\end{align*}
where $L(\beta)$ is the objective function in equation (\ref{appAbs:10}). Then,
\begin{align*}
\lim\limits_{s\rightarrow 0}k(u,s)&=
u'X'Xu-2u'N\bigg(0,\sigma^2 (X'X)\bigg)+\lambda\sum_{i=1}^{m}\bigg(|u_i|I(\beta_i=0)+u_i sign(\beta_i+u_i)\bigg)
,
\end{align*}
where $N$ denotes a normally distributed random variable.
\end{theorem}
\begin{proof}
	Recalling $L(\beta)$ from equation (\ref{appAbs:10}), then
	\begin{align}
	 \lim _{s\rightarrow 0}k(u) &=\lim _{s\rightarrow 0} L(\beta+u)-\lim _{s\rightarrow 0} L(\beta) \nonumber\\
	& = (e-Xu)'(e-Xu)-e'e+\nonumber\\
	&\qquad\qquad\qquad
	\lim\limits_{s\rightarrow 0}\bigg\{2\lambda\sum_{i=1}^{p}
	\beta_i \int_{\frac{\beta_i}{s}}^\frac{\beta_i+u_i}{s}\frac{1}{\sqrt{\pi}}e^{-t^2}dt+
	\lambda \sum_{i=1}^{p}u_i\bigg(2\Phi(\frac{\beta_i+u_i}{s},0,\frac{1}{\sqrt{2}})-1\bigg)
	\bigg\} \nonumber\\
	&=u'X'Xu-2u'N\bigg(0,\sigma^2 (X'X)\bigg)+
	\lim\limits_{s\rightarrow 0}\lambda\sum_{i=1}^{p}
	\bigg(2\beta_i \frac{u_i}{s}\phi(\frac{u_i}{s},0,\frac{1}{\sqrt{2}})+\nonumber\\
	&\qquad\qquad\qquad\qquad\lim\limits_{s\rightarrow 0}\bigg\{ u_i\bigg(2\Phi(\frac{\beta_i+u_i}{s},0,\frac{1}{\sqrt{2}})-1\bigg)\bigg\}\bigg)\nonumber\\
	&=u'X'Xu-2u'N\bigg(0,\sigma^2 (X'X)\bigg) \nonumber\\
	& \qquad\qquad +\lim\limits_{s\rightarrow 0}\lambda\sum_{i=1}^{p}
	\left\lbrace\begin{array}{ll}
	 u_i\bigg(2\Phi(\frac{u_i}{s},0,\frac{1}{\sqrt{2}})-1\bigg) &  \beta_i=0\\
	2\beta_i \frac{u_i}{s}\phi(\frac{u_i}{s},0,\frac{1}{\sqrt{2}})
	+
	u_i\bigg(2\Phi(\frac{\beta_i+u_i}{s},0,\frac{1}{\sqrt{2}})-1\bigg)
	& \beta_i\neq 0
	\end{array}
	\right.\nonumber\\
	&=u'X'Xu-2u'N\bigg(0,\sigma^2 (X'X)\bigg)+\lambda\sum_{i=1}^{p}
	\left\lbrace\begin{array}{ll}
	|u_i| &  \beta_i= 0\\
	u_i &  \beta_i+u_i > 0\\
	-u_i &  \beta_i+u_i < 0
	\end{array} ,
	\right.\nonumber
	\end{align}
	and
	\[k(u)=u'X'Xu-2u'N\bigg(0,\sigma^2 (X'X)\bigg) + \lambda\sum_{i=1}^{p}\bigg(|u_i|I(\beta_i=0)+u_i \rm{sign}(\text{$\beta_i$}+u_i)\bigg).\]
\end{proof}
Theorem (\ref{ch2:theorem1}) shows that  the limit distribution of estimations  under the new penalty is similar  to lasso, see~\citep[Theorem 1]{knight00}, provided $s$ is close enough to zero. That is, the penalization is capable of producing sparse estimations. The theorem however does not provide any optimal value for $s$ to ensure this convergence. In the next theorem we  show that the minimum speed of $s$ that guarantees the convergence of estimations to lasso is $n^{-(1/2+\epsilon)}$ for any $\epsilon>0$.
\begin{theorem}\label{ch2:Theorem 2}
Let $\beta$ be a sparse set of coefficients, $u\in \mathbb{R}^p$, $s_n=s/({n^{1/2+\epsilon}})\rightarrow 0$, $\epsilon>0$, ${\lambda_n }/{\sqrt{n}}\rightarrow \lambda_\circ \geq 0$, and ${X'X}/{n}\rightarrow \Sigma$   where $\Sigma$ is non-singular. Then, $\sqrt{n}(\hat{\beta}_n-\beta){\rightarrow} \arg\min_u k(u)$ where,
\begin{align*}
k(u)=-2u'N(O,\sigma^2 \Sigma) + u' \Sigma u + \lambda_\circ \sum_{i=1}^{p}\{u_i sign(\beta_i)I(\beta_i \neq 0)+|u_i|I(\beta_i=0)\}.
\end{align*}
\end{theorem}
\begin{proof}
	Consider $k_n(u)=L(\beta+\frac{u}{\sqrt{n}})-L(\beta)$. Then
	 \begin{align*}
	k_n(u)& = (e-X\frac{u}{\sqrt{n}})'(e-X\frac{u}{\sqrt{n}})-e'e+\nonumber\\
	& \qquad\qquad\qquad
	2\lambda_n\sum_{i=1}^{p}
	\beta_i \int_{\frac{\beta_i}{s}}^\frac{\beta_i+\frac{u_i}{\sqrt{n}}}{s}\frac{1}{\sqrt{\pi}}e^{-t^2}dt +\lambda_n \sum_{i=1}^{p}\frac{u_i}{\sqrt{n}}\bigg(2\Phi(\frac{\beta_i+\frac{u_i}{\sqrt{n}}}{s},0,\frac{1}{\sqrt{2}})-1\bigg)\nonumber\\
	% % % % % % % % % % %
	&\overset{n\rightarrow \infty}{\rightarrow} u' \Sigma u-2u'N(0,\sigma^2 \Sigma)+\nonumber\\
	& \qquad\qquad\qquad
		\lim\limits_{n\rightarrow \infty}2\lambda_n\sum_{i=1}^{p}
	\beta_i \int_{\frac{\beta_i}{s}}^\frac{\beta_i+\frac{u_i}{\sqrt{n}}}{s}\frac{1}{\sqrt{\pi}}e^{-t^2}dt
	+\lim\limits_{n\rightarrow \infty}\lambda_n \sum_{i=1}^{p}\frac{u_i}{\sqrt{n}}\bigg(2\Phi(\frac{\beta_i+\frac{u_i}{\sqrt{n}}}{s},0,\frac{1}{\sqrt{2}})-1\bigg)\nonumber\\
	&= u' \Sigma u-2u'N(0,\sigma^2 \Sigma)+\nonumber\\
	&\qquad\qquad\qquad
		2\lambda_\circ\sum_{i=1}^{p}
	\beta_i \frac{u_i}{s\sqrt{\pi}}\lim\limits_{n\rightarrow \infty}e^{-(\frac{\beta_i+\frac{u_i}{\sqrt{n}}}{s})^2}
	+\lambda_\circ \sum_{i=1}^{p}{u_i}\bigg(\lim\limits_{n\rightarrow \infty}2\Phi(\frac{\beta_i+\frac{u_i}{\sqrt{n}}}{s},0,\frac{1}{\sqrt{2}})-1\bigg)\nonumber\\
	&= u' \Sigma u-2u'N(0,\sigma^2 \Sigma)\nonumber\\
	&\quad
	+\lim\limits_{n\rightarrow \infty}\left\lbrace
	\begin{array}{ll}
	% FIRST LINE
	 \lambda_\circ \sum_{i=1}^{p}{u_i}\bigg(2\Phi(\frac{\frac{u_i}{\sqrt{n}}}{s},0,\frac{1}{\sqrt{2}})-1\bigg) &  \beta_i\in S_\circ\\
	% SECOND LINE
	2\lambda_\circ\sum_{i=1}^{p}\bigg(
	\beta_i \frac{u_i}{s\sqrt{\pi}}e^{-(\frac{\beta_i+\frac{u_i}{\sqrt{n}}}{s})^2}
	+
	{u_i}\bigg(2\Phi(\frac{\beta_i+\frac{u_i}{\sqrt{n}}}{s},0,\frac{1}{\sqrt{2}})-1\bigg)\bigg)
	 & \beta_i \in S_\circ^c
	\end{array}
	\right.
	,
	%\label{appAbs:22}
	\end{align*}
	%\begin{figure}[H]
	%\centering
	%\includegraphics[width=0.9\linewidth]{"Figures/c2plots/compare_Ubounds"}
	%\caption{Different approximations for the absolute function}
	%\label{fig:compare-Ubounds}
	%\end{figure}
		where $S_\circ$ and $S_\circ^c$ are sets of zero and non-zero coefficients respectively.
\end{proof}
Similar to the derivation of \cite{knight00}, one can show that this results guarantees a sparse estimation of the parameters.
In the proof of  Theorem (\ref{ch2:Theorem 2}), we assumed that $s_n\sqrt{n}\rightarrow 0$. Thus, in practice, if one chooses any $s$ less than $1/\sqrt{n}$, the resulting estimators are similar to lasso. \\

\section{\ {Algorithm for parameter estimation}}\label{sec:dlasso-alg}
The penalised likelihood (\ref{appAbs:10}) is differentiable with respect to $\beta$, so standard optimization routines can be used to find its minimum. These however can be slow. In this section we propose an efficient algorithm, which exploits the differentiability of the dlasso penalty function. To this end, we follow ~\citep{fan01} and define an iterative algorithm as,
\begin{align}\label{algorithm}
\beta^{(k)}= \bigg(X'X +  \Sigma(\beta^{(k-1)},\lambda,s) \bigg)^{-1} X'y,\qquad k=1,2,\ldots
\end{align}
where $\beta^{(0)}$ is an initial estimation for the parameters and $\Sigma(\beta^{(k-1)},\lambda,s)$ is defined by,
\begin{align*}
\Sigma(\beta^{(k-1)},\lambda,s)=\lambda \text{Diag }\bigg[\bigg( 2\Phi(\frac{\beta^{(k-1)}_i}{s},0,\frac{1}{\sqrt{2}})-1 +2\frac{\beta^{(k-1)}_i}{s} \phi(\frac{\beta^{(k-1)}_i}{s},0,\frac{1}{\sqrt{2}}) \bigg)/\beta^{(k-1)}_i,i=1,\ldots,r\bigg].
\end{align*}

In order to derive this, we take the first order Taylor approximation of the dlasso penalty function around $\beta^{(0)}$ given by,
\[\beta (2\Phi(\frac{\beta}{s},0,\frac{1}{\sqrt{2} } )-1)\approx
 \beta^{(0)} (2\Phi(\frac{\beta^{(0)}}{s},0,\frac{1}{\sqrt{2} } )-1) + \bigg(2\Phi(\frac{\beta^{(0)}}{s},0,\frac{1}{\sqrt{2} } )-1
+\frac{2\beta^{(0)}}{s}\phi(\frac{\beta^{(0)}}{s},0,\frac{1}{\sqrt{2} } \bigg)(\beta-\beta^{(0)}).\]
Note that the differentiability of dlasso means that we do not need to resort to local quadratic approximations as in \cite{fan01}.
Given $\beta^{}\approx\beta^{(0)}$, we can now rewrite this as
\begin{align}
\beta (2\Phi(\frac{\beta}{s},0,\frac{1}{\sqrt{2} } )-1)&\approx
 \beta^{(0)} (2\Phi(\frac{\beta^{(0)}}{s},0,\frac{1}{\sqrt{2} } )-1) +\nonumber\\ &\frac{1}{\beta^{(0)}}\bigg(2\Phi(\frac{\beta^{(0)}}{s},0,\frac{1}{\sqrt{2} } )-1
+\frac{2\beta^{(0)}}{s}\phi(\frac{\beta^{(0)}}{s},0,\frac{1}{\sqrt{2} } \bigg)(\beta^2-\beta^{(0)^2}).
\label{eq:alg}
\end{align}

Substituting (\ref{eq:alg}) into the penalised likelihood (\ref{appAbs:10}) results in
\begin{align*}
(y-X\beta)'(y-X\beta) & + \lambda \sum_{j=1}^{p} \bigg[
 \beta_j^{(0)} (2\Phi(\frac{\beta_j^{(0)}}{s},0,\frac{1}{\sqrt{2} } )-1) +\nonumber\\
& \frac{1}{\beta_j^{(0)}}\bigg(2\Phi(\frac{\beta_j^{(0)}}{s},0,\frac{1}{\sqrt{2} } )-1
+\frac{2\beta_j^{(0)}}{s}\phi(\frac{\beta_j^{(0)}}{s},0,\frac{1}{\sqrt{2} } \bigg)(\beta_j^2-\beta_j^{(0)^2})\bigg],
\end{align*}
the optimum of which can be found efficiently by iteratively computing the ridge regression as in (\ref{algorithm}).

This algorithm has been implemented in the R package {\tt DLASSO}, which is freely available from CRAN, \url{http://CRAN.R-project.org/package=DLASSO}. The package allows also to select the tuning parameters $s$ and $\lambda$ by common model selection criteria, such as Akaike Information Criterion (AIC), Bayesian Information Criterion (BIC) or Generalized Cross-Validation (GCV).

\section{Simulation study} \label{sec:simulation}
We have performed a simulation study to assess the performance of the new penalty in a regression context, $y=X\beta+\sigma e$, $e\sim N(0,1)$. Similar to  \cite{zou05} we design three simulation scenarios.
\begin{enumerate}
\item[Scenario 1]: \textit{Standard}.  We set $\beta=(3,1.5,0,0,2,0,0,0)$, simulate the predictors from a multivariate normal distribution with mean zero and correlation $\mathbb{C}or(X_i,X_j)=0.5^{|i-j|}$ and generate the response with $\sigma^2=3$.
\item[Scenario 2]: \textit{Small $\beta$s.} Same as the first scenario except that $\beta_j=0.5, j=1,2,\ldots,8$.
\item[Scenario 3]: \textit{Correlated predictors}. We consider $p=15$ and divide the coefficients into three groups, $\beta^{(1)}=c(1,2,3,4,5)$, $\beta^{(2)}=c(0.5,0.5,0.5,0.5,0.5)$ and $\beta^{(3)}=c(0,0,0,0,0)$. We consider a high correlation of $0.9$ amongst each pair of the first five covariates. Similarly,  we assume a  correlation of $0.5$ in the second group, whereas we assume no dependency in the third group.  Finally, we set $\sigma^2=15$.
\end{enumerate}
For all three scenarios, we generate $50$ datasets containing $240$ observations: $40$ observations are assigned to the training set and the remaining are assigned to the test set. The penalty parameters are tuned on the training set using 10-fold Cross-Validation (CV) on the mean squared error.

We compare the following models: dlasso (with fixed $s=0.01$), dlasso (with both $s$ and $\lambda$ tuned), lasso, ridge, , elastic net (enet), SCAD and Ordinary Least Squares regression (OLS). For lasso and elastic net, we use the R package {\tt msgps} \citep{hirose12}, for SCAD we use the R package {\tt ncvreg} \citep{breheny11}, for ridge we use the R package {\tt glmnet} \citep{friedman10}. For the dlasso methods, we use the R package {\tt DLASSO}. In the first scenario, we also consider the log-approximation of \cite{schmidt07} as in Equation \ref{appAbs:4} (with $s=0.01$). However, this penalty turned out to be rather unstable in the optimization, so we did not include it in the other scenarios.

Figure 3 shows the results. For each scenario, we plot the distribution (over the 50 iterations) of the mean squared error on the test set, defined by $\dfrac{\sum_{i=1}^{200} y_i -\hat y_i}{200}$ and the mean squared error of the estimated parameters, defined by $(\hat\beta-\beta)'S_X(\hat\beta-\beta)$, with $\beta$ and $\hat\beta$ denoting the true and estimated values of the parameters, respectively. Among all the penalties, the first four plotted are differentiable at zero (OLS, ridge, dlasso with s=0.01 and general dlasso), the other ones are popular penalties in the regularized regression literature. Overall, the dlasso penalty performs better or the same as existing penalties and is superior to ridge, which is the only alternative differentiable penalty for regularized problems, and to SCAD, which is the only alternative non-convex penalty.

%\begin{center}
%FIGURE 3 ABOUT HERE
%\end{center}
\begin{figure}[ht!]
\centering
\includegraphics[scale=0.6]{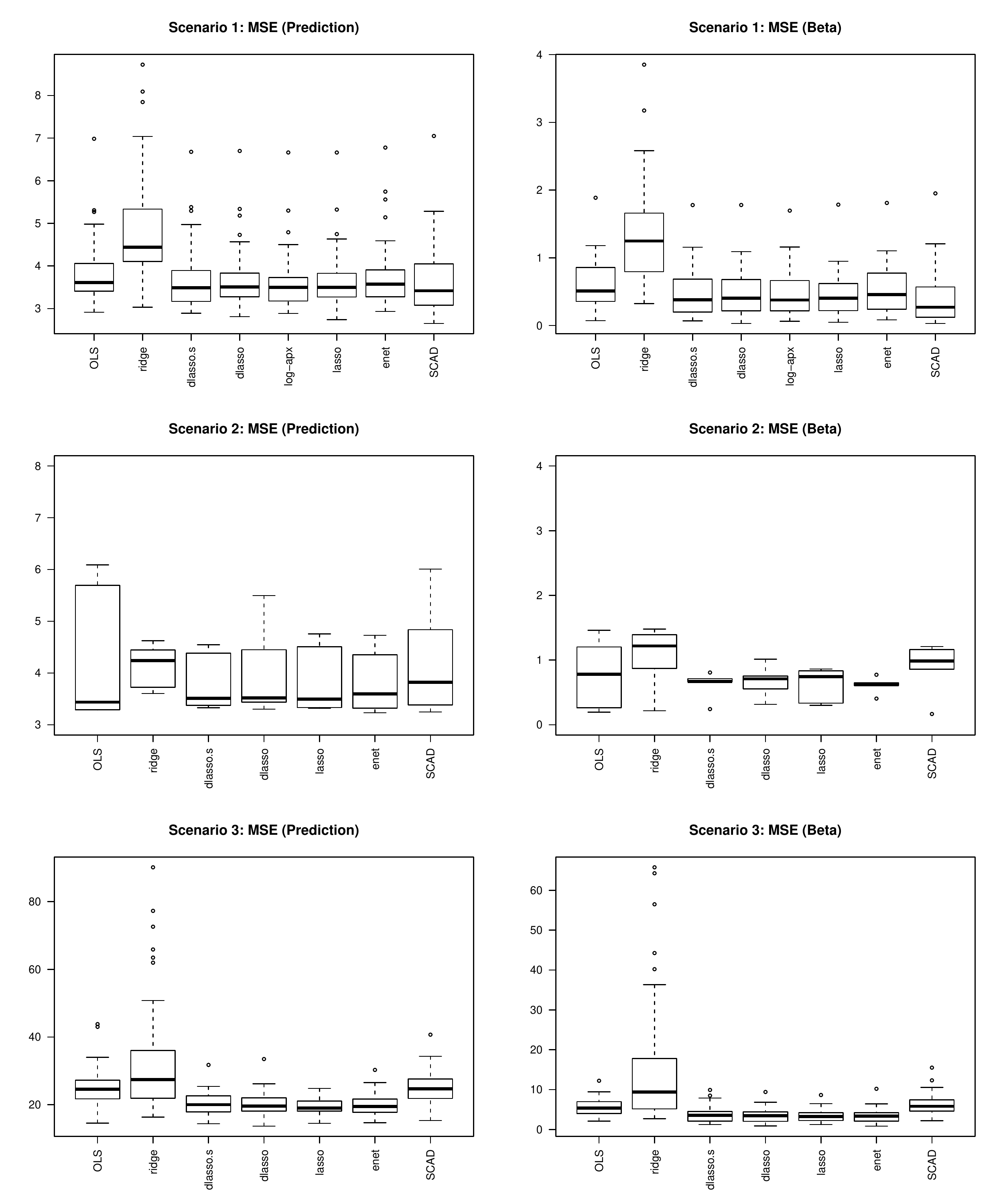}
\caption{Simulation study comparing dlasso with existing methods on three scenarios. Two versions of dlasso are considered: one where both $s$ and $\lambda$ are tuned by 10-fold CV (referred to as {\tt dlasso}), and the other where $s$ is fixed at 0.01, i.e. close to an $l_1$ penalty, and $\lambda$ tuned by CV (referred to as {\tt dlasso.s}). The plots report the Mean Squared Error (MSE) of the prediction on the test set (left) and of the estimated regression parameters (right).}
\label{fig:simulation}
\end{figure}

\section{Prostate cancer example} \label{sec:realdata}
 We consider the prostate data by~\cite{stamey89}, previously analysed by \cite{zou05} using regularized regression methods. The objective of the analysis is to investigate the correlation between the level of prostate specific antigen (lpsa) and a number of clinical measurements in 97  men who were about to receive a radical prostatectomy. There are eight covariates: log cancer volume (lcanvol), log prostate weight (lweight), log benign prostatic hyperplasia amount (lbph), log capsular penetration (lcp), age,  Gleason score (gleason), percentage Gleason scores 4 or 5 (pgg45),  seminal vesicle invasion (svi). All covariates are normalized to have zero mean and unit variance and the response to have zero mean.

Lasso, ridge, SCAD, OLS, elastic-net and dlasso are applied to the data. BIC is used to select all tuning parameters. For dlasso, we also consider the case of $s$ fixed to 1, where the results are expected to be similar to ridge, and $s$ fixed to 100, where we expect a solution similar to OLS. The results in Table 1 show a similar performance of dlasso compared with lasso and elastic net. All three models select the same five predictors and are superior to SCAD in terms of AIC and BIC. The comparison with ridge and OLS confirms our expectations. Finally, dlasso shows a better BIC compared to OLS, probably due to the effect of a small amount of regularization still present for $s=100$.

%\begin{center}
%TABLE 1 ABOUT HERE
%\end{center}

\begin{table}[ht!]
	\caption{Comparison of lasso, ridge, SCAD, OLS, elastic-net and dlasso for $s=0.001$ (BIC optimal), $s=1$ (ridge), $s=100$ (OLS) on the prostate dataset. All tuning parameters are selected by BIC. The methods are compared based on AIC, BIC and sparsity.}
	\centering
		\begin{tabular}{llcccl}		
		 & 	 & 	 & 	 & 	 & 	\\
		\textbf{Method} & 	\textbf{Precision} & 	\textbf{AIC} & 	\textbf{BIC} & 	\textbf{df} & 	\textbf{Significant Variables} \\
		\hline dlasso & 	s=0.001 & 	207.6 & 	216.1 & 	5 & 	lcavol, ibph, lweight, pgg45,svi \\
		\hline lasso & 	- & 	206.7 & 	215.3 & 	5 & 	lcavol, ibph, lweight, pgg45,svi\\
		\hline elastic net & 	$\alpha$=0.001 & 	206.8 & 	215.3 & 	5 & 	lcavol, ibph, lweight, pgg45,svi\\
		\hline SCAD & 	- & 	214.6 & 	231 & 	4 & 	lcavol, ibph, lweight,svi\\
		\hline & 	 & 	 & 	 & 	 & 	\\
		\multicolumn{6}{c}{\textbf{Ridge} } 	\\
		\hline dlasso & 	1 & 	207.4 & 	227 & 	8 & 	all variables\\
		\hline ridge & 	- & 	207.1 & 	226 & 	8 & 	all variables\\
		\hline & 	 & 	 & 	 & 	 & 	\\
		\multicolumn{6}{c}{\textbf{OLS} } 	\\
		\hline dlasso & 	100 & 	204 & 	228 & 	8 & 	all variables\\
		\hline OLS & 	- & 	202 & 	233 & 	8 & 	all variables\\
		\hline
		\end{tabular} 
\end{table}

% Similar to~\citep{zou05}, we divide the dataset into a training set consisting of $67$ observations and a test set consisting of $30$ observations. We center all predictors and response in the training set to mean zero and further scale the predictors to variance 1. The same centering and scaling factors are used on the test set. We use 10-fold Cross-Validation (CV) on the training set to select the tuning parameter and to estimate the regression parameters. We then compare different models based on the prediction mean-squared error (MSE) on the test data. Figure(\ref{fig:prostate}) reports the results for a number of models considered and across 100 splits of training and test. The figures do not show much difference at the level of MSE (Figure(\ref{fig:prostate}), left), with the exception of ridge regression which under-performs the other methods, but there is some difference in the degree of sparsity of the selected methods (Figure(\ref{fig:prostate}), right), with lasso and dlasso (with fixed $s$=0.01) returning the most sparse solutions.
%\begin{figure}[p!]
%	\centering
%	\includegraphics[width=0.48\linewidth]{Prostate-MSEPrediction}
%	\includegraphics[width=0.48\linewidth]{Prostate-Sparsity}
%	\caption{Analysis of Prostate data. Distribution of MSE on test set for 100 train-test splits and distribution of the number of non-zero estimated coefficients for a range of models.}
%		\label{fig:prostate}
%\end{figure}

\section{Conclusions} \label{sec:conclusion}
In this paper, we have proposed a novel penalty term that is capable of producing similar results to other well-known penalty functions in the context of regularized regression. One key difference, however, is that this new penalty is differentiable. This opens up the possibility of using it in many contexts where differentiability plays a key role. For example, a differentiable objective function could lead to more efficient implementations of parameter estimation procedures for certain models or to improved model selection criteria by a more accurate estimation of the bias term. These aspects will be investigated in future work.

%\begin{table}[p!]
%	\caption{Comparison of lasso, ridge, SCAD, OLS, elastic-net and dlasso for $s=0.001$ (BIC optimal), $s=1$ (ridge),$s=100$ (OLS) on the prostate dataset. All tuning parameters are selected by BIC. The methods are compared based on AIC, BIC and sparsity.}
%	\centering
%	\begin{tabular}{lccccc}
%		 \multicolumn{6}{c}{\textbf{Lasso}}\\
%		\hline \textbf{Method} 				& \textbf{Precision} 	  & \textbf{AIC} &	\textbf{BIC} &   \textbf{df}  & \textbf{Significant variables}\\
%		\hline dlasso 					& s=0.001   & 	207.6 	& 216.1 		  &   5  & lcavol, ibph, lweight, pgg45,svi\\
%		\hline lasso 					&  - 	  & 	206.7 	& 215.3 		  &   5  & lcavol, ibph, lweight, pgg45,svi\\
%		\hline elastic-net				&  $\alpha$=0.001  & 	206.8 	& 215.3 		  &   5  & lcavol, ibph, lweight, pgg45,svi\\
%		\hline SCAD						&  -	  & 	214.6 	& 231.0 		  &   4  & lcavol, ibph, lweight, svi\\
%		\hline \multicolumn{6}{c}{}\\
%		 \multicolumn{6}{c}{\textbf{Ridge}}\\
%		\hline dlasso 					& 1	      & 	207.4 	& 227	 	&   8  & all variables\\
%		\hline ridge 					&  -	  & 	207.1 	& 226 		&   8  & all variables \\
%		\hline \multicolumn{6}{c}{}\\
%		\multicolumn{6}{c}{\textbf{OLS}}\\
%		\hline dlasso					& 100	  & 	204 	& 228		 		  &   8  & all variables \\
%		\hline OLS	 					& -		  & 	202 	& 233		 		  &   8  & all variables \\
%		\hline \multicolumn{6}{c}{}\\
%	\end{tabular}
%	\label{tb:prostate}
%\end{table}
\bibliographystyle{plain}
\bibliography{dlasso}
\end{document}